\documentclass{article}
\usepackage[T1]{fontenc} 
\usepackage[utf8]{inputenc} 
\usepackage{ismir,amsmath,amssymb,cite,url}
\usepackage{graphicx}
\usepackage{color}

\usepackage{siunitx}

\newcommand{\new}[1]{#1}

\title{Online Symbolic Music Alignment with Offline Reinforcement Learning}

\oneauthor
 {Silvan David Peter} {Institute of Computational Perception, Johannes Kepler University Linz, Austria \\ {\tt silvan.peter@jku.at}}

\def\authorname{S. Peter}

\begin{document}

\maketitle
\begin{abstract}

Symbolic Music Alignment is the process of matching performed MIDI notes to corresponding score notes.
In this paper, we introduce a reinforcement learning (RL)-based \textit{online} symbolic music alignment technique. 
The RL agent --- an attention-based neural network --- iteratively estimates the current score position from local score and performance contexts.
For this symbolic alignment task, environment states can be sampled exhaustively and the reward is dense, rendering a formulation as a simplified offline RL problem straightforward.
We evaluate the trained agent in three ways.
First, in its capacity to identify correct score positions for sampled test contexts; second, as the core technique of a complete algorithm for symbolic online note-wise alignment; and finally, as a real-time symbolic score follower.
We further investigate the pitch-based score and performance representations used as the agent's inputs.
To this end, we develop a second model, a two-step Dynamic Time Warping (DTW)-based \textit{offline} alignment algorithm leveraging the same input representation. 
The proposed model outperforms a state-of-the-art reference model of offline symbolic music alignment.

\end{abstract}

\section{Introduction}\label{sec:introduction}

Music alignment refers to matching at least two different versions of the same musical material.

In this paper, we address symbolic music alignment, for our purposes defined as models that match individual notes of a performance recorded as MIDI file to individual notes of a score encoded as MusicXML file.

Alignment procedures can be separated into online or offline classes.
If the alignment procedure is carried out with access to the full versions of the musical material, we refer to it as offline alignment. 
Conversely, if 
one version is only known up to the point currently to be matched, we refer to it as online.

We introduce a reinforcement learning (RL)-based \textit{online} symbolic music alignment technique. 
It aligns symbolically encoded music or, more specifically, MIDI performances to their corresponding MusicXML scores by matching individual notes of each version.
The RL agent -- a small attention-based neural network -- is trained to iteratively predict the current score position from limited score and past performance contexts.
The current performance note and estimated score position are then processed to compute a symbolic note-wise matching.

RL terminology introduces another
online versus offline differentiation.
RL is termed online if the agent learns from data created by the agent's interaction with its environment during training.
In our case, we use offline RL, that is, the agent is trained using a dataset of exhaustively sampled environment states and associated rewards, effectively turning agent training into a supervised learning problem.
Once trained in an offline fashion, the agent can be used in online alignment.

The agent processes a purely \emph{pitch-based representation} and timing information is only incorporated in a post-processing step.
Before addressing the online problem, we investigate the same separation of pitch and time processing in an offline setting:
we develop a two-step (first pitch, then time) Dynamic Time Warping (DTW) offline model and evaluate it against the state of the art in note-wise alignment in Section~\ref{sec:offline}.
The subsequent Section~\ref{sec:online} addresses the RL-based online model reusing the input setup. 

The rest of this paper is thus structured as follows:
Section~\ref{sec:rel_work} introduces related work.
Section~\ref{sec:offline} discusses offline symbolic music alignment. 
We develop as well as evaluate an offline symbolic music alignment algorithm based on two different applications of DTW, first on pitch information, then on onset times.
Starting from these results, section~\ref{sec:online} introduces a formulation of online alignment as reinforcement learning problem.
In particular, we train an agent's value function in an offline setting.
In section~\ref{sec:online_eval_full} we evaluate the trained agent in three ways: as a standalone score onset identification model, as an online symbolic alignment model (where the aim is the production of correct note-wise alignments), and in a score following scenario (where the aim is the precise temporal tracking the current score position).
Finally, Section~\ref{sec:discussion} concludes the paper with a critical appraisal of our models as well as recommendations for future research.

\section{Related Work}\label{sec:rel_work}

Symbolic music alignment has been a popular research area for many years.
We begin our review of related work with online symbolic music alignment, then we progress to offline symbolic music alignment, general music alignment, and finally applications of reinforcement learning.

Most often, online models have been presented in the context of score following, where the principal aim is to identify the current score position.
Dannenberg~\cite{dannenberg1984line} and Vercoe~\cite{vercoe1984synthetic} pioneered this area of research in the mid 1980s.
Recent works commonly use Dynamic Bayesian Networks to track the performance~\cite{Nakamura2014MergedOutputHM, Raphael2009OrchestralAF, nakamura2015autoregressive,maezawa2011polyphonic}.
Recently, Cancino-Chac{\'o}n et al.~compared both Hidden Markov Models (HMM) and On-Line Time Warping (OLTW) techniques~\cite{cancino2023accompanion}.
We use their OLTW model as comparison baseline for our online alignment technique.
This model processes inputs represented as piano rolls, as is common for OLTW and DTW applications to symbolic music alignment in general.

The offline setting has seen more recent work~\cite{gingras2011improved,chen2014improved, Nakamura2014MergedOutputHM,Nakamura:2015,nakamura2014outer,nakamura2017performance, nakamura2015autoregressive}.
Symbolic music alignment methods often perform very well, with error rates rarely exceeding 10\%.
Consequently, much recent work focused on the rare, indeterminate, or asynchronous events that make the errors difficult to identify and fix.
Ornaments are one source of such events~\cite{Nakamura:2015}.
Another is left-right hand asynchrony in piano performance as discussed in Nakamura et al.~\cite{Nakamura2014MergedOutputHM}.
Arbitrary skips and repeats further present a very difficult challenge for most algorithms, especially when runtime considerations are important~\cite{nakamura2014outer}.
This series of articles by Nakamura et al.~\cite{Nakamura:2015,Nakamura2014MergedOutputHM,nakamura2014outer} culminated in one of the most widely used automatic score-performance alignment tools and the current state of the art (SOTA)~\cite{nakamura2017performance}.
We use this model as reference for the evaluation of our offline algorithm.

Although beyond the scope of this article, no introduction of music alignment is complete without the mention of the large body of work concerning alignment of non-symbolic music formats, in particular audio. 
Wang~\cite{Wang:thesis}, Arzt~\cite{arzt:2016}, and chapter three in Müller~\cite{Mueller:2015} present introductory discussions of audio alignment.
As in our offline approach, applications of (non-standard) DTW are central to audio alignment~\cite{praetzlich:2016,muller2021sync,tralie2020exact,ewert2009high}. 
Audio score following is commonly computed using On-Line Time Warping and variants of Hidden Markov Models~\cite{dixon2005line,cont2008antescofo,raphael2010music,duan2011state,Arzt2015RealTimeMT}.

To the best of our knowledge, Dorfer et al.~\cite{matthias_dorfer_2018_1492535} \new{(later expanded upon by Henkel et al.~\cite{henkel2019score})} are the only prior application of RL in a music alignment task, namely online audio to sheet music image alignment.
For a general introduction to RL, we refer the reader to Sutton and Barto~\cite{sutton2018reinforcement}, for a discussion of the merits and disadvantages of offline RL to Levine et al.~\cite{levine2020offline}.

\section{Offline Symbolic Music Alignment}\label{sec:offline}

In this section, we introduce an offline symbolic music alignment based on two different DTW steps as well as an intermediate cleanup step.
We close the section with an evaluation of our model against a state-of-the-art reference.

Symbolic music alignment produces \textit{note alignments}, i.e., it matches individual notes of a performance recorded as MIDI file to individual notes of a score encoded as MusicXML file.
Three types of note alignments exist: a match is tuple of a performance note and a score note, a deletion is a score note not played, and an insertion is a performed note not notated.

Our proposed offline algorithm consists of the following steps:
First, the performance and score are aligned using DTW on a purely pitch-based representation (Section~\ref{sec:pitchseq}).
Then, remaining gaps are filled by complete sequences of a single pitch.
Finally, individual notes are aligned using an application of DTW on their onset times  (Section~\ref{sec:onset_dtw}).

\begin{figure}
 \centerline{\framebox{
 \includegraphics[width=0.97\columnwidth]{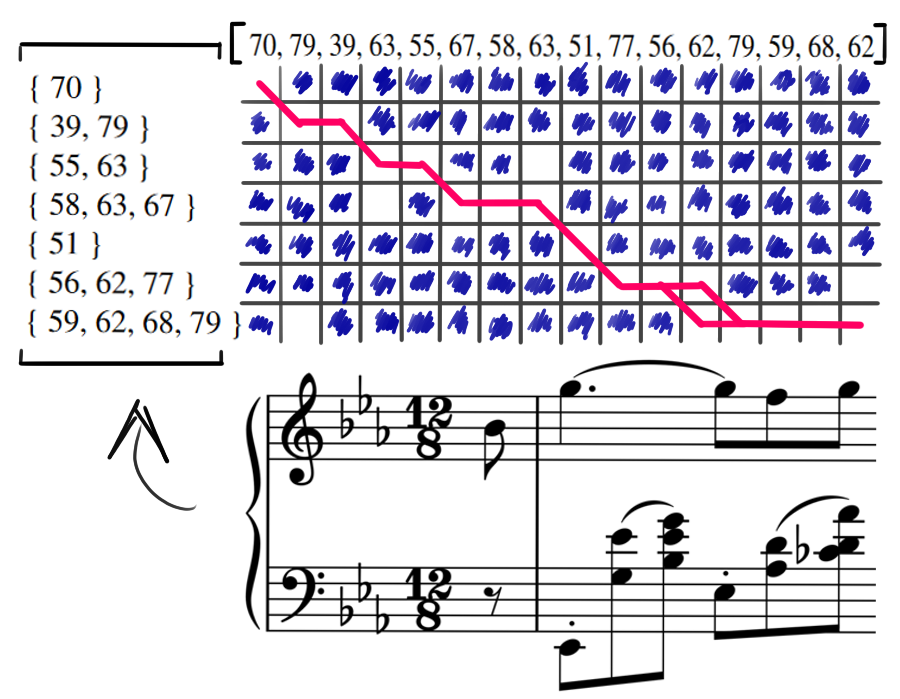}}}
 \caption{First half measure of Chopin Op. 9 No. 2 (bottom score), encoded as pitch set sequence (left) and warped to its performance, encoded as sequence of pitches as played (top).
 The matrix shows the corresponding pairwise distance (shaded is distance of 1, see equation~\ref{eq:dtw}), red lines indicate equivalent optimal warping paths.}
 \label{fig:dtw}
\end{figure}

\subsection{Pitch Sequence Warping}\label{sec:pitchseq}

In this approach,
we align sequences of performance notes, encoded as integer pitches $ p_t \in \mathbb{I} :=\{1...88\}$, with sequences of score onset notes, encoded as sets of integer pitches $s_t \in \mathcal{P}(\mathbb{I}) \setminus \{\varnothing \}$, \new{with $\mathcal{P}(\mathbb{I})$ denoting the power set of the set $\mathbb{I}$}.
Since these sequence elements are of different types --- integers and sets of integers --- no standard local distance metric can be used.
Instead we opt for a non-symmetric inclusion metric, with some abuse of the term.

\begin{equation}\label{eq:dtw}
    m(p_{t},s_{t}) =
\left\{
	\begin{array}{ll}
		 0 & \mbox{if } p_{t}\in s_{t} \\
		 1 & \mbox{else } 
	\end{array}
\right.
\end{equation}

Having defined the metric in Equation~\ref{eq:dtw}, two standard DTW paths are computed, one forward and one backward, i.e., using inverted sequences.
Figure~\ref{fig:dtw} shows the encoding as well as examplary DTW paths computed from cumulative pairwise distances.
While the optimal DTW distance is unique, the DTW paths are not necessarily so.
In our case, such ambiguity is often introduced by repeated pitches in neighboring score onsets, see e.g.,~the \new{two adjacent (left, stacked vertically) notes of} pitch D4/62 in Figure ~\ref{fig:dtw}.
To pinpoint non-robust path segments, we use the backward DTW path.
Wherever the forward and backward paths disagree, they effectively bracket ambiguous parts from both sides, and we exclude all bracketed notes from the path.

These excluded segments are then processed using a simple heuristic: 
The notes in bracketed segments are separated by pitch.
If two pitch-wise sequences with matching number of notes (in the performance and the score) are found, they are aligned and the result is added to the path.
\new{If no matching sequence is found, the path is linearly interpolated.}
We finally compute a mapping from score time to performance time from this merged and cleaned path.

\subsection{Onset Sequence Warping}\label{sec:onset_dtw}

The next goal is to derive note-wise alignments from the approximate score to performance mapping computed in Section~\ref{sec:pitchseq}.
To this end, the performance and score are split into pitch-wise sequences for each pitch occurring in the union of score and performance pitches.
The approximate score-time-to-performance-time mapping computed in the previous step is used to project all score onsets \new{(in beats)} to performance time points \new{(in seconds)}.
The last step aligns the performance onset sequence with the score onset sequence, now mapped to the same space.

This alignment is computed by a DTW path between the onset sequences, this time using a simple $L_1$ metric, and for each non-unique alignment, keeping the tuple with the lowest distance.
A threshold of maximal distance is further built-in (and set to 5 seconds) to avoid spurious alignment of unrelated deletions and insertions.

\subsection{Offline Model Evaluation}\label{sec:offline_eval}

To test the full offline model (first pitch DTW~\ref{sec:pitchseq}, then onset DTW~\ref{sec:onset_dtw}), we compute alignments on four datasets of high-quality note-wise alignment piano music.
The datasets are the Vienna 4x22 Dataset~\cite{vienna4x22}, four excerpts performed by 22 performers each; the Zeilinger Dataset~\cite{CancinoChacon:2017ht}, nine piano sonatas by Ludwig van Beethoven performed by Clemens Zeilinger; the Magaloff Dataset~\cite{magaloff}, the near complete solo piano works by Frederic Chopin performed by Nikita Magaloff; and the Batik Dataset~\cite{widmer2003playing}, twelve piano sonatas by Wolfgang Amadeus Mozart performed by Roland Batik. 
We compare our model against the reference by Nakamura et al.~\cite{nakamura2017performance}, post-processed to produce the same output format we employ.

To compare produced alignments to ground truth ones, we have to define a metric.
Recall that note alignments consist of three types: matches (tuples of performance and score notes), deletions (unplayed score notes), and insertions (unnotated performed notes).
We use an F-score metric for matches:
A predicted match is counted as a true positive (TP) only if the same notes are matched in the ground truth alignment.
A false positive (FP) is a predicted note label that isn't in the ground truth, a false negative (FN) is a ground truth note label that isn't predicted.
The F-score is defined as the harmonic mean of precision (TP / (TP + FP)) and recall (TP / (TP + FN)).

\begin{table}[ht]
    \centering
    \begin{tabular}{@{}l|r|r@{}}
  \hline
Dataset & DTW Offline & Nakamura  \\ 
  \hline
Magaloff & 98.4 $\pm$ 0.9 \%	& 97.8 $\pm$ 1.4 \%	\\
Zeilinger & 99.3 $\pm$ 0.9 \%	& 98.8 $\pm$ 1.2 \%	\\
Batik & 99.4 $\pm$ 0.7 \%	& 98.5 $\pm$ 2.1 \%	\\ 
Vienna 4x22 & 99.8 $\pm$ 0.4 \%	& 99.5 $\pm$ 0.5 \%	\\
Combined & 99.0 $\pm$ 1.0 \%	& 98.5 $\pm$ 1.5 \%		\\ 
  \hline
\end{tabular}
    \caption{Dataset-wise averaged F-scores and standard deviations of each model.}
    \label{Tab:offline}
\end{table}

Table~\ref{Tab:offline} shows dataset-wise and globally averaged F-scores for matches.
Our proposed model outperforms the reference on each dataset.
A two-sided sign test on performance-wise rankings shows significantly ($\alpha=0.01$) higher performance for our proposed model on all datasets except the Vienna 4x22 Dataset. 
On the Vienna 4x22 dataset, the models reach the same F-score of 1.0 for 38 performances and our proposed model has higher F-scores for the remaining 50 performances.

\section{Online Alignment Agent}\label{sec:online}

Having established the effectiveness of the \new{separation into} pitch-based and \new{time-based} input representations in the offline setting, we now introduce a formulation of RL-based online alignment.
We continue with the model and training setup used to approximate the agent's value function.

Reinforcement learning is formalized as Markov Decision Process (MDP).
An MDP consists of the following components: 
a state space $\mathcal{S}$, an action space $\mathcal{A}$, an index set $\mathcal{T}$, a reward function $\mathcal{R}$, transition probabilities $\mathcal{P}$, and a discount factor $\mathcal{\gamma}$.

An agent is placed in an environment and perceives this environment and itself as being in a possible state $S_t \in \mathcal{S} (t \in \mathcal{T})$.
The agent now takes an action $A_t \in \mathcal{A}$ and receives a reward $R_{t+1}$ as well as a new state $S_{t+1} \in \mathcal{S}$.
Repeating this process iteratively yields a sequence of states, actions, and rewards, called an episode: 
$ S_t, A_t, R_{t+1}, S_{t+1}, A_{t+1}, R_{t+2}, S_{t+2}, A_{t+2}, ...$
It is now the agent's task to infer actions from states that maximize long-term reward.
Before we look at our formulation of this optimization problem, we discuss the online alignment's state and action space in more detail.

\subsection{Alignment as Reinforcement Learning}

The state information $S_t$ comprises both the current score context as well as the most recent past performance.
Specifically, the score context is represented as a window of the pitch set sequence introduced in Section~\ref{sec:pitchseq}.
The window centers the last predicted score onset position
and spans seven score onsets to the past as well as eight score onsets to the future for a total windowed sequence of 16 pitch sets.
The performance context is only derived from past performance notes to enable real-time application.
It consists of the eight most recent notes in the performance pitch sequence.
Whenever less \new{score or performance} context is available, e.g., at the very beginning or end of a piece, the windows are shortened accordingly.

At each state $S_t$ the agent aims to match the most recent performance note to its most likely score onset.
\new{There are 16 actions at $S_t$; select one score onset as matching onset position.}
Having decided on a next score onset, the agent receives a reward $R_{t+1}$ which is set to one if the score onset is correctly aligned, zero otherwise.
The environment transition probabilities $\mathcal{P}$ determine a new state $S_{t+1}$:
\new{the performance context of the new state is determined by an actual performance, i.e., the agent is presented a new state based on the estimated next position in the score and a new incoming performance note.}

\subsection{Simplified Deep Q-learning}\label{sec:rl_online}

The agent's behavior is captured by its policy $\pi(A | S)$, the distribution of actions taken by the agent in state $S$.
Although it is possible to optimize the policy directly, we instead adapt a value-function-based formulation, or more specifically, deep Q-learning~\cite{watkins1992q,mnih2015human}.
Q-learning aims to optimize a state-action value function $Q: \mathcal{S}\times\mathcal{A} \longrightarrow \mathbb{R} $, an estimate of the expected cumulative discounted future reward, also called return, given a state and an action.
In deep Q-learning the value function $Q(S,A,\theta)$ contains trainable parameters $\theta$ which are fitted to the experienced reward distribution.
A typical optimization loss $l$ looks like this:

\begin{equation}
    l =  ( R_{t+1} + \gamma \max_{A_{t+1}} Q(S_{t+1},A_{t+1},\theta) - Q(S_{t},A_{t},\theta) )^2 
 \end{equation}

where the discount factor $\gamma \in [0,1]$ determines the weighting of future rewards.
For the alignment case we can make several simplifications.
We opt for a completely myopic agent, i.e., $\gamma = 0$.
The argument for this is that the optimal action to take for each incoming performance note is determined by the correct score onset which can in turn be specified by the immediate reward.
>For a discussion of the implications of this modeling choice see section~\ref{sec:discussion}.
Setting $\gamma = 0$ removes the value function at subsequent states from the loss.
Using that $ \mathcal{R} = \{0,1\}$, we make a second reformulation and replace this squared error loss by a binary classification: 
For each state-action tuple $(S,A)$ the agent predicts the probabilities of the reward being 1 or 0, optimized with a cross-entropy loss.
Note that this formulation still optimizes a state-action value function $Q$ and not a policy $\pi(A | S)$, i.e., the probabilities of rewards are computed for each possible action (that is, per score onset) and do not sum to one over all actions.
There are several ways of deriving a policy from a value function; two are discussed in section~\ref{sec:full_model}.

\begin{figure}
 \centerline{\framebox{
 \includegraphics[width=0.97\columnwidth]{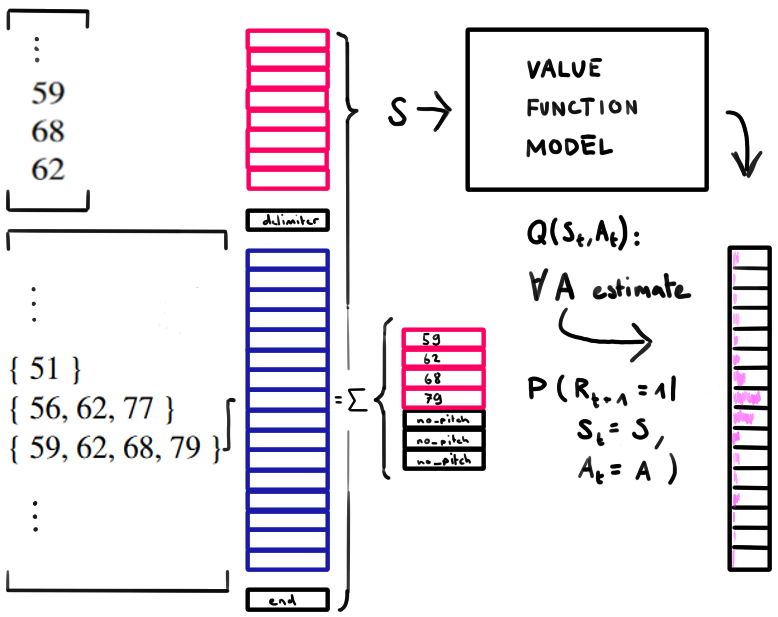}}}
 \caption{Setup of the value function model: states are encoded as contiguous token sequence of past performance (red) and current score (blue) contexts. Pitch set embeddings are summed over individual pitch embeddings. The model is set up as token classifier as each score onset in the context corresponds to a possible action (= "select this onset as next score onset") and is classified according to its expected reward class. The vector on the right shows the reward probability for each action (pink).}
 \label{fig:value_model}
\end{figure}

\subsection{Value Function Model}

To approximate $Q(S,A,\theta)$, we use an attention-based Transformer Neural Network.
The input of the network consists of a sequence of tokens encoding the performance, a delimiter token, the score, and an ending token.
We encode 88 pitches of the piano keyboard, adding extra tokens for "no\_pitch", "delimiter", and "end" in a 64-dimensional embedding space.
The performance pitches are straightforward to embed, however, the score onset pitch sets require more processing.
For our data, more than 99\% of score onsets can be represented with pitch sets with no more than seven different pitches, we thus limit our pitch sets to this length (with a subset of seven taken randomly at onsets with more pitches).
Pitch sets with fewer than seven pitches are filled up with a pitch corresponding to the "no\_pitch" token.
To create score onset embedding with no more than 64 dimensions independent of the number of pitches at any onset, the pitch set tokens are summed up.
Figure~\ref{fig:value_model} shows the setup of the value function model.
The model is set up with eight heads, and six layers, layer normalization, and a single feedforward head for binary classification, making for a total of 157250 parameters.

\subsection{Training}
The training is set up as token classification problem; i.e., for each token in the sequence, the probability of receiving a reward is estimated.
The \new{aligned piano} datasets from Section~\ref{sec:offline_eval} are used again.
For our offline RL setting, a dataset of states is created before training.
We extract local score and performance contexts from the aligned data, shifting the performance window such that \new{true next} score onsets fall from leftmost (current position minus seven) to rightmost (current position plus eight) to cover the possible states exhaustively.
\new{During training, batches of states are sampled randomly, not in sequence.}
To aid generalization, we further augment the data by random pitch shifting \new{of all notes in the state} within +/- one octave.
We use an ADAM optimizer with a learning rate with warm-up followed by square root decay.
The batch size is set to 8192, the models are trained for 50 epochs.

\subsection{Online Models}\label{sec:full_model}

\begin{figure}
 \centerline{\framebox{
 \includegraphics[width=0.97\columnwidth]{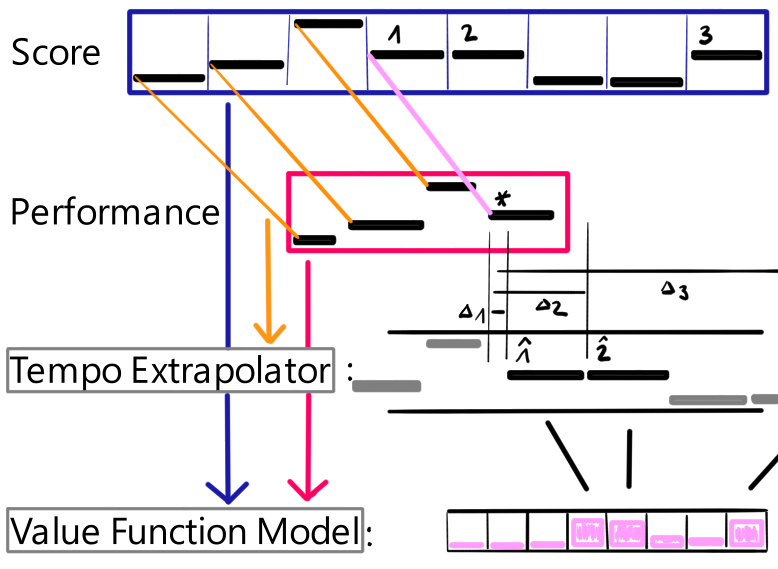}}}
 \caption{
 Schematic overview of the Online Alignment Model with a monophonic piece and 8 onset context.
 Score (blue, top) and performance (red) contexts are inputs to the Value Function Model which outputs value estimates (pink, bottom).
 The top three onsets (1,2,3) are passed to a tempo extrapolator, along with existing alignments (yellow).
 The tempo extrapolator predicts three onsets ($\hat{1},\hat{2}$) for the the candidate onsets.
 The one with lowest distance ($\Delta_1$) to the newest performance note (*) is aligned (pink).
 }
 \label{fig:alignmnet_model}
 \vspace*{-5pt}
\end{figure}

Using a trained value function model, we derive two complete models.
First, a simple score follower model ("Greedy Agent Model") that outputs \new{only greedily} estimated score onsets for incoming performance notes.
Second, a note-level alignment model ("Online Alignment Model") that produces both note alignments as well as \new{an estimate of the} next score onset for the score following setting.
The Greedy Agent Model consists of an agent following a greedy policy based on the trained value function model, i.e., an agent picking the action $A$ with maximal estimated reward $Q(S,A)$.

For the Online Alignment Model a few additional steps are taken.
See Figure~\ref{fig:alignmnet_model} for an overview of the alignment loop.
In this model, an action is selected from the top three value estimates for a given score and performance context.
To pick one of these three actions, onset time information is incorporated.
A simple local tempo estimator approximates an expected performed onset time for each of the three possible score onsets using linear extrapolation of beat periods computed from previously aligned notes.
This process takes on the role of the second onset-wise DTW step in the offline model (see Section~\ref{sec:onset_dtw}): 
to match notes that are closest together according to an approximate score-to-performance mapping.

There are two further heuristics worth mentioning.
If the current performed pitch is not available at any of the three highest ranking score positions, the performed note is counted as an insertion, and the
current score position is unchanged.
Furthermore, we decrease the number of calls made to the agent in a real-time setting by directly aligning pitches that are trivially missing at the current score onset.

\section{Online Evaluation}\label{sec:online_eval_full}

In the following, we evaluate the agent and the proposed online alignment model.
In section~\ref{sec:agent_eval}, we address a greedy agent's capacity to identify correct score positions for sampled test contexts.
In section~\ref{sec:online_eval}, the Online Alignment Model is evaluated with respect to correct note-wise alignment. 
Finally, we use both the Greedy Agent Model as well as  the Online Alignment Model as real-time symbolic score followers in an experiment in section~\ref{sec:sf_eval}.

\subsection{Agent Evaluation}\label{sec:agent_eval}

\begin{table}[ht]
    \centering
    \begin{tabular}{l|l|l}
\hline
  Top0  &  Top1 & Top2 \\
\hline
94.5 $\pm$ 0.8 \% & 96.6 $\pm$ 0.5 \% & 97.6 $\pm$ 0.4 \% \\
\hline
\end{tabular} 
    \caption{Average topK score onset hit rate and standard deviation across the five test folds.}
    \label{tab:agent_eval}

\end{table} 

For direct
value function evaluation, we assume a greedy policy for each testing state $S$.
That is, the agent picks the action $A$ with the highest estimated value $Q(S,A)$.
We evaluate this action (= chosen score onset) via the distance from the ground truth score onset.
Specifically, we compute three metrics: the number of states where this action corresponds exactly to the true score onset ("Top0"), the number of times this action picks a score onset in the neighborhood of $\pm$ one score onset of the true location ("Top1"), and the number of times this action picks a score onset in the neighborhood of $\pm$ two score onsets of the true location ("Top2"), each normalized by the total number of test states.
We use five-fold cross-validation on the same combined datasets used in section~\ref{sec:offline_eval}, and report mean and standard deviation values across testing folds.
The fold splitting is carried out piece-wise with roughly the same number of score onsets in each fold.

Table~\ref{tab:agent_eval} shows the results.
Greedy action selects the correct score onset with more than 94 \% probability on unseen pieces.
Furthermore, for more than half of the remaining errant actions, the greedy action is not further than two onsets from the correct one.

\subsection{Online Note-wise Alignment}\label{sec:online_eval}

\begin{table}[ht]
    \centering
    \begin{tabular}{@{}l|r|r|r@{}}
  \hline
Piece & OAM & DTW Offline & Nakamura \\ 
  \hline
B. Op.~53 3rd.~m.& 99.0  \%	& 99.4 \%	& 98.2 \%	\\
C. Op.~9 No.~1 & 97.6 \%	& 98.4 \%	& 98.8 \%	\\
C. Op.~9 No.~2 & 97.4 \%	& 99.1 \%	& 97.6 \%	\\ 
C. Op.~10 No.~11 &  90.3  \%	& 96.3 \%	& 94.3 \%	\\
C. Op.~60  & 95.1 \%	& 97.9 \%	& 94.7 \%	\\ 
  \hline
\end{tabular}
    \caption{Piece-wise F-scores of each model. OAM = Online Alignment Model, DTW Offline = model of section ~\ref{sec:offline_eval}, Nakamura = reference SOTA model~\cite{nakamura2017performance}.}
    \label{tab:eval_online}
\end{table}

To evaluate the Online Alignment Model's performance, we perform alignments for five selected performances: Nocturnes Op.~9 No.~1 and 2, Etude Op.~10 No.~11, Nocturne Op.~15 No.~2, the Barcarole Op.~60 by F. Chopin, and the third movement of the Sonata Op.~53 (Waldstein) by L. v. Beethoven.
The value function model used in this section was trained on all data except these five pieces for 100 epochs, the rest of the training and model setting remains the same.
The same metrics of section~\ref{sec:offline_eval} apply, namely the F-score of retrieved matched note tuples.
For comparison we also add piece-wise F-scores of our proposed offline model as well as the model by Nakamura et al.
Table~\ref{tab:eval_online} shows the F-scores of note alignments.
As expected, the proposed online alignment performs worse than offline methods, albeit with small difference.
Notably, all models show the lowest performance on Chopin's Op. 10 No. 11.

\subsection{Score Following}\label{sec:sf_eval}

In the score following setting, the core metric is the accurate  prediction of the current position.
We thus compute asynchrony values in milliseconds which give the absolute time between any onset in a performance and the onset in the same performance that corresponds to the estimated score onset.
The data used for this experiment consists of the same five pieces used in the previous section~\ref{sec:online_eval} with the same value function model training.
Three models are compared in this setting:
The Online Alignment Model (OAM) previously evaluated in terms of note alignment F-scores in Table~\ref{tab:eval_online}, the Greedy Agent Model (GAM), and an On-Line Time Warping (OLTW) Model.
\new{This latter OLTW} model performed best in a recent music score following comparison by Cancino-Chac{\'o}n et al.~\cite{cancino2023accompanion} and is added as a reference.
\new{However, this model does not predict note alignments, hence a comparison in terms of note alignment F-score as in Section~\ref{sec:online_eval} is not possible.}

\begin{table}[t]
    \centering
    \begin{tabular}{l|r|r|r|r}
\hline
Model  &  Async  &  $\leq25\si{\milli\second}$ & $\leq50\si{\milli\second}$  &  $\leq100\si{\milli\second}$   \\
\hline
OLTW & 60.6 \si{\milli\second}& 38.0 \% & 63.3 \% & 86.7 \% \\
GAM & 36.0 \si{\milli\second} & 89.0 \% & 91.4 \% & 94.6 \%  \\
OAM & 15.7 \si{\milli\second} & 91.4 \% & 93.8 \% & 96.6 \%    \\
\hline
\end{tabular} 
    \caption{Asynchrony of the models in score follower setting. Column "Async" presents the median asynchrony. Columns 3, 4, 5 present the percentage of onset estimates with lower asynchrony than 25ms, 50ms and 100ms, respectively.}
    \label{tab:sf}
\end{table}

Both the GAM and the OAM outperform the reference model in all metrics in Table~\ref{tab:sf}.
Most of the lower performance of the GAM is due to the fact that for Chopin's Op. 10 No. 11, this agent loses track of the performance close to the end when a full measure is deleted.
All subsequent performance notes are estimated very wrongly.
The online alignment model on the other hand follows all test performances robustly until the end.

\section{Discussion and Conclusion}\label{sec:discussion}

In this paper, we introduce two models, an offline alignment model based on dual DTW steps, and an online alignment model based on an RL agent trained in an offline fashion.
Both models perform competitively; with the offline model surpassing the relevant state of the art.

In the setup of the RL training we made some simplifications that warrant further discussion.
Specifically, we set the discount factor $\gamma$ to zero and train using a dataset of sampled states.
In section~\ref{sec:rl_online}, we claim that the optimal action for each step is determined by the correct score onset.
While this is true for the states in the dataset and if optimality is defined by accuracy in note-wise alignment, it might not be for out-of-distribution states or if the focus of the agent is shifted to robustness, i.e., following the entire performance even at the cost of some misaligned notes.

For offline RL, a crucial issue is distributional drift~\cite{levine2020offline}; i.e., the fact that the agent learns from states that follow a different distribution that the states it would encounter in an online setting.
Even though we can sample the state space exhaustively for the training set, out-of-distribution states are expected in test sets consisting of previously unseen pieces and performances.
Furthermore, the relative frequency of training samples does not necessarily correspond to the states an online agent is likely to see during training, where all target locations have the same frequency.
Specifically, for an agent that already learned to predict the score onset with some accuracy, the targets at the limits of the context are going to be less frequent than the center ones.
In other words, a non-myopic online agent is likely to behave more conservatively, avoiding large skips as they do not occur that frequently in actual performances.

On the other hand, the offline RL formulation successfully leverages prior knowledge about the task and --- more importantly --- stabilizes the gradient, rendering the training of a complex value function approximator feasible.
Future work includes shifting this trade-off back towards online RL, for example with online RL training after initial offline training.

The RL agent learns to align purely on pitch information.
Including onset or even duration information is likely to increase the accuracy of following at the cost of requiring a more expressive model which in turn affects inference speed --- a hard bottleneck for real-time application.
In fact, running the value estimation for every incoming performance note (such as the score follower "GAM" in Table \ref{tab:sf}) uses up to a minute of computation time for the 7273 notes in the performance of Beethoven's Op. 53 Mvt. 3 (Roughly 10 \si{\milli\second} per note).
A further increase is liable to affect real-time score following in fast passages.

In terms of post-processing steps, both our offline and online models are comparatively crude, making little use of score information such as ornaments.
As Nakamura et al.~\cite{nakamura2017performance} correctly remark, their post-processing step is in principle able to improve upon any prior more error-prone alignment.
Further research is needed to know whether the offline model can be improved in this way.

To conclude, we developed and evaluated two models of symbolic music alignment which both outperform relevant prior work.
To the best of our knowledge, this RL-based online alignment model is one of the first applications of not only trainable but effectively \textit{trained} models to symbolic music alignment.

\section{Reproducibility}\label{sec:reproducibility}
Our alignment models are available at:
\url{https://github.com/sildater/parangonar}\\

\section{Acknowledgements}

This work is supported by the European Research Council (ERC) under the EU’s Horizon 2020 research \& innovation programme, grant agreement No.~101019375 (“Whither Music?”).

\bibliography{ISMIR2023_template}

\end{document}